\documentclass[multphys,vecphys]{svmult}

\usepackage{makeidx}         % allows index generation
\usepackage{graphicx}        % standard LaTeX graphics tool
\usepackage{multicol}        % used for the two-column index
\usepackage[bottom]{footmisc}% places footnotes at page bottom
\makeindex             % used for the subject index
\begin{document}

\title*{Dissolution of Globular Clusters}
\author{Holger Baumgardt}
\institute{Argelander Institute for Astronomy, Auf dem H\"ugel 71, 53121 Bonn
\texttt{holger@astro.uni-bonn.de}}
\maketitle

\section{Introduction}
\label{sec:1}

Globular clusters are among the oldest objects in galaxies, and understanding the details
of their formation and evolution can bring valuable insight into the early history of
galaxies. Until the late 1970s, globular clusters were thought of to be relatively static 
stellar systems, a view which was supported by the fact that most observed density profiles
of globular clusters can be fitted with equilibrium models like e.g. King (1966) profiles.
This view has changed significantly over the last twenty years. On the observational side,
the evidence for differences in the stellar mass-functions of globular clusters
(Piotto, Cool \& King 1997, de Marchi et al.\ 1999), which are believed to be at least
partly the result of their dynamical evolution, and
the discovery of extratidal stars surrounding globular clusters (Grillmair et al.\ 1995,
Odenkirchen et al.\ 2003) are strong indications for the ongoing dynamical evolution and 
dissolution of globular clusters.

On the theoretical side, $N$-body simulations of star cluster evolution have become increasingly
sophisticated, due to both progresses in simulation techniques 
(e.g. Mikkola \& Aarseth 1993, Aarseth 1999) and the development of the GRAPE series of special purpose
computers (Sugimoto et al.\ 1990, Makino et al.\ 2003), which allows to simulate the evolution of star
clusters with increasingly larger particle numbers.

This review summarises the current knowledge about the dissolution of star 
clusters and discusses the implications of star cluster dissolution for the evolution of the mass 
function of star cluster systems in galaxies.

\section{Dissolution mechanisms}
\label{sec:2}

Star clusters evolve due to a number of dissolution mechanisms, the most important of which are:

\begin{itemize}
\item[(1)] Primordial gas loss
\item[(2)] Stellar evolution
\item[(3)] Relaxation
\item[(4)] External tidal perturbations
\end{itemize}

The importance of the different processes changes as a star cluster ages and 
depends also on the position of the cluster in its parent galaxy. 
The combined effect of all dissolution mechanisms can dramatically alter the properties of
star cluster systems, so it is important to understand them.

\subsection{Primordial gas loss}

Star formation is typically less than 40\% efficient and the gas not turned into stars is
lost within a few $10^5$ to $10^6$ Myrs due to stellar winds from massive stars or supernova explosions. $N$-body simulations 
have shown that the loss of the primordial gas can easily cause star clusters to lose a large fraction of their 
stars (Goodwin 1997, Kroupa et al.\ 2001, Boily \& Kroupa 2003) or unbind them completely. Together with the loss of a large
mass fraction within a short timescale, star clusters also undergo significant expansion. Connected to the 
problem of primordial gas loss 
is the question whether star clusters form in virial equilibrium, since, in addition to gravity, molecular
clouds are also held together by magnetic fields and the pressure from the ambient gaseous medium.
Since these forces do not act on stars, some clusters might already be unbound at birth and disperse
within a few crossing times. As a result, a significant fraction of clusters might not survive the first
10 Myrs (''infant mortality problem'', Lada \& Lada 2003).

These considerations are supported by observations which show that a large fraction of clusters dissolve at 
an early stage. Fall et al. (2005) for example found that in the Antennae galaxies the number of clusters
decreases strongly with cluster age and that the median age of clusters is only $10^7$ yrs,
which they interpreted as evidence for rapid cluster disruption. Similarly, only a small
fraction of star-forming embedded clusters in the Milky Way evolve to become open clusters (Lada \& Lada 2003),
implying that the majority of clusters must dissolve within a few Myrs. 

\subsection{Stellar evolution}

For clusters which survive the early evolutionary stages, the next dissolution mechanism is stellar
evolution. For a standard stellar IMF like for example the Kroupa (2001) IMF, about 30\% of the mass of a star 
cluster is lost due to the 
stellar evolution of the member stars within a few Gyrs (Baumgardt \& Makino 2003) and the fraction can be significantly
higher for star clusters starting with a top-heavy IMF. Mass-loss from stellar evolution causes the
clusters to expand while at the same time decreasing the tidal radius of star clusters. Although the
effect is less strong than the loss of the primordial gas since the mass loss happens
on a longer timescale, $N$-body simulations have shown that the effect is strong enough to 
unbind low-concentration clusters surrounded by an external tidal field.

Fukushige \& Heggie (1995) for example found that if the stellar IMF follows a power-law with 
index $\alpha=-1.5$ between $0.4 < m < 15$ M$_\odot$, clusters less concentrated than King $W_0=7$ models 
were easily disrupted by
stellar evolution mass-loss. For Salpeter-like IMFs, clusters had to have an initial concentration
$W_0 \ge 5$ to be stable against disruption. Similar results were also found by Giersz (2001) and
Joshii et al. (2001) in Monte-Carlo simulations.
\begin{figure}[t]
\centering
\includegraphics[height=8cm]{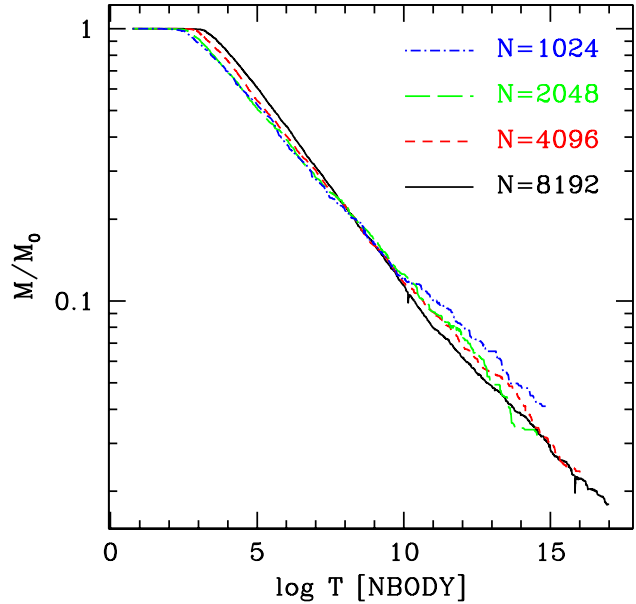}
\caption{Dissolution of isolated clusters starting with different initial particle numbers $N$. Irrespective of $N$, complete
dissolution takes about $10^{15}$ $N$-body times since relaxation causes a strong cluster expansion,
which slows down the overall evolution (from Baumgardt et al.\ 2002).}
\label{fig:1}       % Give a unique label
\end{figure}

\subsection{Relaxation}

Relaxation arises due to mutual encounters between stars in a globular cluster, and causes a slow
drift of stars in energy space. Dynamical processes like the segregation of heavy mass stars into the 
cluster center or the core collapse of star clusters are driven by relaxation. For a star cluster containing 
$N$ stars with mean mass $<\!m\!>$ and half-mass radius $r_h$, the relaxation time at $r_h$ is given by 
(Spitzer 1987, eq.\ 2-63):
\begin{equation}
t_{rh} = 0.138 \frac{\sqrt{N} \;\; r_h^{3/2}}{\sqrt{<\!m\!>} \; \sqrt{G} \; \ln(\gamma N)} \; \; ,
\end{equation}
where $G$ is the constant of gravity and $\gamma$ the Coulomb logarithm. 

Due to mutual encounters between cluster stars, stars can also gain enough energy to leave 
the cluster completely, which causes a slow evaporation of the whole
cluster. $N$-body simulations (Baumgardt et al.\ 2002 and Fig.\ 1) have shown that for {\it isolated} clusters
this process is inefficient in dissolving star clusters since even low-mass clusters would need
of order $10^{15}$ initial crossing times to completely evaporate, i.e. they would not dissolve           
within the lifetime of the universe. The
reason is that relaxation is dominated by the cumulative effect of many distant encounters, so stars
change their energies smoothly and do not jump around in energy space. As the energy of a star approaches
$E \rightarrow 0$, it inevitably moves through the outskirts of the cluster for most of the time where the 
stellar density is low and the star has only few encounters with other stars. As a result, relaxation alone 
causes clusters to expand but does not dissolve them.

\subsection{External tidal fields}

Star clusters are usually not isolated but move in the gravitational field of their parent galaxy. The
external galaxy can influence the evolution of a star cluster in two ways: A constant tidal field,
which arises for example if clusters move on circular orbits through axisymmetric potentials,
confines the cluster stars into a certain volume around the cluster centre, 
outside of which the stars are unbound to the cluster. This prevents clusters from expanding indefinitely
and accelerates the escape of stars since the energy necessary to escape from the cluster is
lowered. 

Variable external fields arise if clusters move on elliptic orbits or through a galactic disc.
In the first case they
experience disruptive tidal shocks since stars are accelerated away from the cluster centre, 
in the latter case the shocks are compressive. In both cases the
internal energy of the clusters is increased. 

\begin{figure}[t]
\centering
\includegraphics[height=3cm]{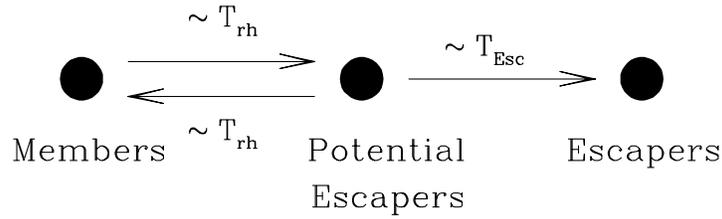}
\caption{Escape of stars from clusters in circular orbits. Bound members are scattered on relaxation time
scales to become potential escapers with $E_* > E_C$. Potential escapers either escape or lose energy and
become bound members again.}
\label{fig:2}       % Give a unique label
\end{figure}

For star clusters moving in circular orbits, stars with an energy $E_*$ only slightly higher than
the critical energy $E_C$ needed for escape can escape only through small apertures around the lagrangian
points L1 and L2 connecting the centre of the galaxy with the centre of the star cluster.
Fukushige \& Heggie (1997) have shown that this leads to a scaling of the average time needed
to escape according to $T_{Esc} \propto 1/(E_*-E_C)^2$, i.e. for small energy differences the escape time can
become of the order of the relaxation time or even larger. This leads to a complication of the whole
escape process, as illustrated in Fig.\ 2: Bound members are scattered on a relaxation time to become
potential escapers with $E_* > E_C$. Potential escapers can either escape or are scattered to lower energies
and become bound members again.

Baumgardt (2001) has shown that this influences the scaling of the lifetimes of star clusters with the
number of cluster stars or the total mass. While the lifetimes of single-mass clusters surrounded with a tidal boundary 
scale with the relaxation time, clusters moving in circular orbits through their parent galaxy show a scaling 
of their lifetimes according to $T_{rh}^{0.75}$. If lifetimes of star clusters are estimated by scaling the 
results of low-$N$ models to higher particle numbers, this causes a reduction of the lifetimes of globular clusters
by a factor of a few. Interestingly, a similar slow increase of the lifetime of star clusters with the particle number 
was also found in observational studies of open cluster systems (Boutloukos \& Lamers 2003, Lamers et al.\ 2005).

\section{Evolution of realistic clusters}

Since the pioneering study by Chernoff \& Weinberg (1990), a number of papers have studied the evolution of multi-mass 
star clusters evolving under the combined effects of stellar evolution, relaxation and an external tidal
field (e.g. Gnedin \& Ostriker 1997, Vesperini \& Heggie 1997, Kroupa, Aarseth \& Hurley 2001). In the following 
I will concentrate on the
results of Baumgardt \& Makino (2003), who have so far performed the largest set of $N$-body calculations of the 
evolution of multi-mass star clusters in external tidal fields. Their clusters all started with Kroupa (2001) IMFs, but 
varying initial particle 
numbers, orbital types and density profiles. 

Fig.\ 3 summarises their results for the dissolution times.
Independent of orbital type and initial cluster concentration, the lifetimes always scaled with the
relaxation time as $T^x_{rh}$ where $x \approx 0.7$. Clusters starting from King models with a larger central
concentration (open circles) show a slightly steeper scaling of the lifetimes. Interestingly, the exponent does 
not change if one goes from circular to elliptic orbits (triangles), indicating that although tidal shocks
help in removing stars, the general picture of stellar escape is not changed significantly. Star clusters moving
at smaller galactocentric distances $R_G$ have smaller lifetimes since the tidal field is stronger and confines
the clusters to smaller volumes. Baumgardt \& Makino (2003) summarised their results for the lifetimes 
with the following formula:
\begin{equation}
\frac{T_{Diss}}{[Myr]} \; = \; \beta \; \left( \frac{N}{ln(\gamma \, N)} \right)^x \;
  \frac{R_G}{[kpc]} \; \left( \frac{V_G}{220\; km/sec} \right)^{-1} (1-\epsilon)
\end{equation}
where $\beta=1.91$, $\gamma = 0.02$ and $x = 0.75$ are constants, $V_G$ is the circular velocity of the
external galaxy and $\epsilon$ the orbital eccentricity of the cluster. This formula did fit the
$N$-body results for star clusters surviving for a Hubble time to within 10\%. It slightly overpredicted the 
lifetimes of star clusters dissolving in less than 1 Gyr since such clusters contain massive stars for a larger fraction 
of their lifetime, which reduces their relaxation times.
\begin{figure}[t]
\centering
\includegraphics[height=10cm]{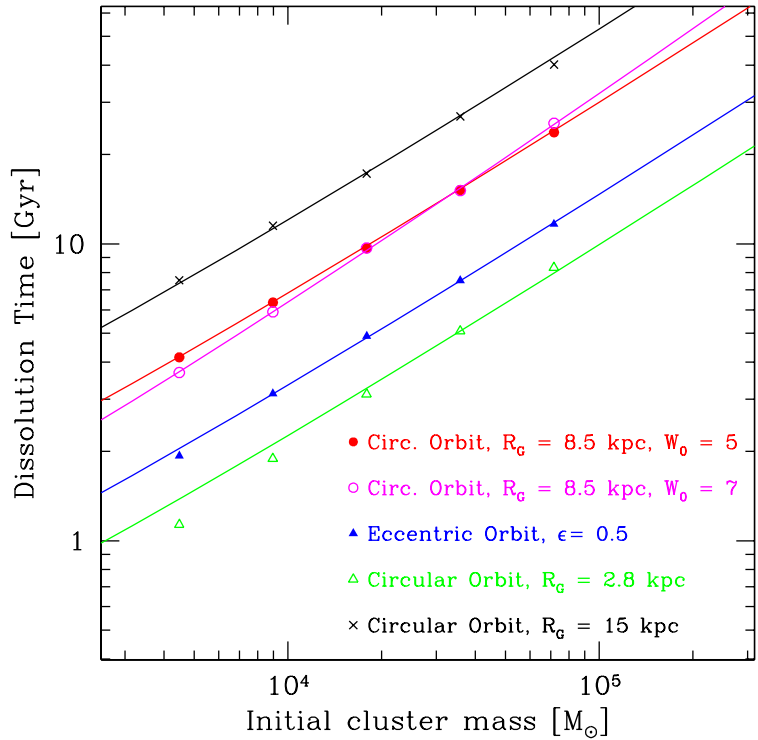}
\caption{Lifetimes of star clusters in dependence of their mass for clusters moving in different
orbits and starting from different initial King models. In all cases, the lifetimes show a scaling with the 
mass close to $T_{Diss} \sim M^{0.75}$.}
\label{fig:3}       % Give a unique label
\end{figure}

\section{Evolution of globular cluster systems}

The globular cluster system of the Milky Way as we observe it today is characterised by a Gaussian distribution in 
absolute magnitudes, with mean $M_V = -7.4$ and scatter $\sigma_M = 1.15$, similar to what is observed for globular 
clusters in other galaxies. In contrast, young massive clusters in interacting and starburst galaxies follow
power-law distributions over mass $N(M) \sim M^{-\beta}$, with slopes close to $\beta = 2$,
so the question arises whether the globular cluster system of the Milky Way started with a similar mass-function
and has lost the low-mass clusters due to dissolution. Stars lost from these clusters could nowadays form the halo 
field stars.

In order to study this question, I have performed a number of Monte-Carlo simulations in which clusters were assumed to 
start with a power-law IMF $N \sim M_C^{-\beta}$ with power-law index $\beta=2.0$ and followed a radial distribution in the 
galaxy according to $\rho \sim R_G^{-\alpha}$ with $\alpha=4.5$ and core radius $R_C=1$ kpc. The galaxy was modelled 
as an isothermal sphere with a circular velocity of $V_G=200$ km/sec and the cluster system was set-up such that 
the clusters had an isotropic velocity dispersion at all radii. The lifetimes derived by Baumgardt \& Makino (2003)
and given in eq. 2 were used to dissolve clusters.

In addition to the dissolution mechanisms already considered by Baumgardt \& Makino (2003), the simulations also included 
dynamical
friction and disc shocks. Dynamical friction was modeled as a steady shrinking of the cluster orbit according to eq. 7-25 of 
Binney \& Tremaine (1987), while disc shocks were included according to Spitzer \& Chevalier (1973), assuming that a fractional
increase of the cluster energy by $\Delta E/E_C$ corresponds to a fractional mass loss of the same amount.
\begin{figure}[t]
\centering
\includegraphics[height=5.8cm]{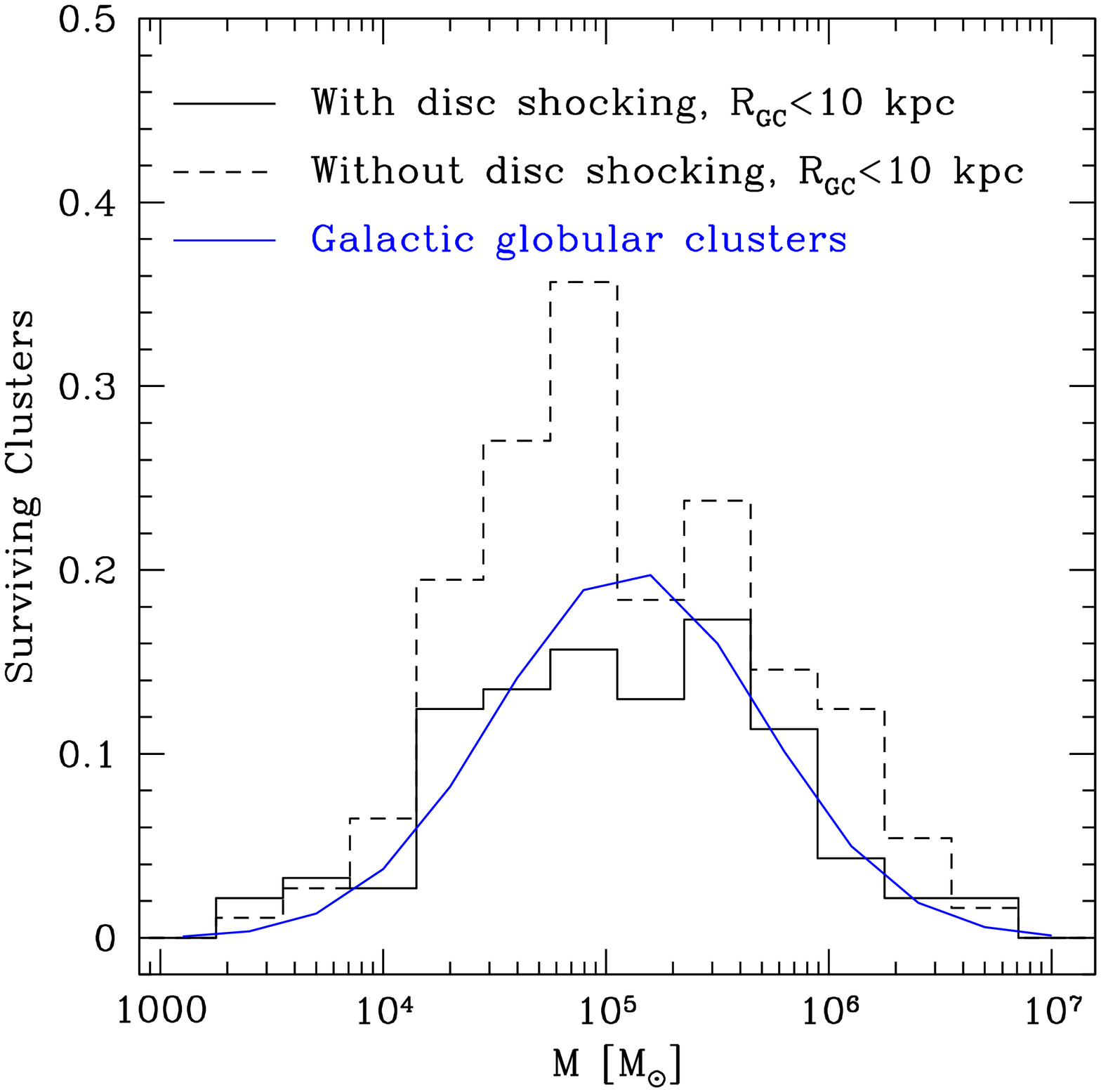}
\includegraphics[height=5.8cm]{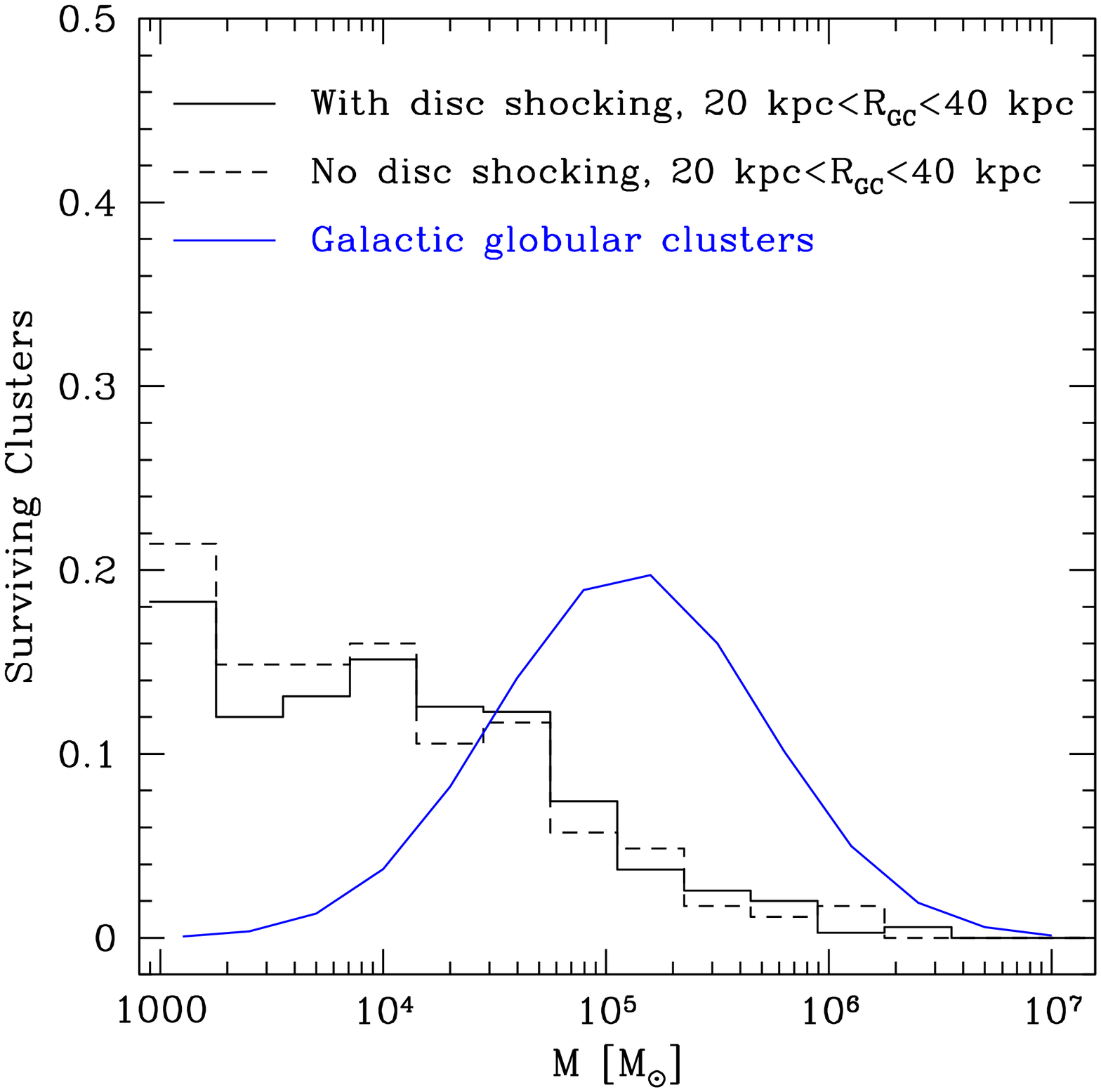}
\caption{Evolution of the mass function of galactic globular clusters for inner ($R_G<10$ kpc, left panel) and
and outer (20 kpc$<R_G<$ 40 kpc, right panel) clusters. While inside 10 kpc, power-law IMFs can be turned into gaussians,
the mass-function of outer clusters is increasing towards the lowest masses and is in contradiction with the 
observed MF.}
\label{fig:4}       % Give a unique label
\end{figure}

Fig.\ 4 shows the resulting evolution of the mass function of the galactic globular cluster system. 
Inside $R_G=10$ kpc, the dissolution mechanisms are strong enough to evolve an initial power-law MF into a 
bell-shaped MF. The mean and dispersion of the surviving clusters agree rather well
with the observed MF of galactic globular clusters. The situation changes if one considers clusters
at larger galactocentric radii. For outer clusters the tidal field is much weaker, meaning less destruction
of the cluster system. As a result, the MF of surviving clusters is still increasing towards the lowest 
masses considered,
in contrast with what is observed for the galactic globulars. This agrees qualitatively with results derived
by Vesperini (1998) and Parmentier \& Gilmore (2005) but is in contrast to what Fall \& Zhang (2001) found for 
the outer clusters. Further research is necessary to constrain the starting condition of the galactic GC system.

%%%%%%%%%%%%%%%%%%%%%%%%%%%%%%%%%%%%%%%%%%%%%%%%%%%%%%%%%%%%%%%%%%%%%%  }

%%%%%%%%%%%%%%%%%%%%%%%%%%%%%%%%%%%%%%%%%%%%%%%%%%%%%%%%%%%%%%%%%%%%%%

\printindex
\end{document}